\documentclass[12pt]{article}

\begin{document}
\thispagestyle{empty}
\noindent\
\\
\\
\\
\begin{center}
\large \bf Composite Weak Bosons, Leptons and Quarks

\end{center}
\hfill
 \vspace*{1cm}
\noindent
\begin{center}
{\bf Harald Fritzsch}\\

University of Munich\\
Faculty of Physics,\\
Arnold Sommerfeld Center for Theoretical Physics,
\vspace*{0.5cm}
\end{center}

\begin{abstract}
The  weak bosons consist of two fermions, bound by a new confining gauge force. The mass scale of this new interaction is determined. At energies below 0.5 TeV the standard electroweak theory is valid. A neutral isoscalar weak boson $X$ must exist - its mass is less than 1 TeV. It will decay mainly into quark and lepton pairs and into two or three weak bosons. Above the mass of 1 TeV one finds excitations of the weak bosons, which mainly decay into pairs of weak bosons. Leptons and quarks consist of a fermion and a scalar. Pairs of leptons and pairs of quarks form resonances at very high energy.\\
\end{abstract}

\newpage

Many years ago the $\rho$ -mesons were assumed to be elementary gauge bosons. Today we know that this assumption is not correct. The $\rho$ -mesons are quark-antiquark bound states, and the masses of the $\rho$ -mesons are due to the field energies of the quarks and gluons inside the mesons, as the masses of the nucleons.\\

In this paper the weak bosons, the leptons and the quarks are assumed to be composite particles. They have a finite size - the present limit on the size of the electron, the muon and of the light quarks is about $10^{-17}$ cm. The constituents of the weak bosons and of the leptons and quarks are bound by a new confining gauge interaction. We denote the constituents as "haplons" - the Greek translation of "simple" is "haplos" - and the new confining gauge theory as QHD. The QHD mass scale is given by a mass parameter $\Lambda_h$. This mass scale determines the sizes and the masses of the weak bosons. A theory of this type was proposed in 1981 ( ref.(1), see also ref.(2-6)).\\

Two types of lefthanded spin 1/2 haplons are needed as constituents of the weak bosons, denoted by $\alpha$ and $\beta$. The doublet h of the weak isospin group $SU(2)$ is given by the two lefthanded haplon fields:\\

\begin{equation} h = \left( \begin{array}{l} \alpha\\ \beta\\ \end{array} \right) \ \end{equation}\\

Their electric charges are  in units of e:\\

\begin{equation}
h = \left( \begin{array}{l}
+\frac{1}{2}\\
-\frac{1}{2}\\
\end{array} \right) \ \end{equation}\\

The three weak bosons have the following internal structure:\\

\begin{eqnarray}
W^+ & = & \overline{\beta} \alpha \; , \nonumber \\
W^- & = & \overline{\alpha} \beta \; , \nonumber \\
W^3 & = & \frac{1}{\sqrt{2}} \left( \overline{\alpha} \alpha -
\overline{\beta} \beta \right) \; .
\end{eqnarray}\\

We expect that in a composite model the structure of the spectral functions of the weak currents is similar to the structure of the spectral functions in hadronic physics. At energies of the order of the QCD mass scale the spectral functions of the hadronic currents are dominated by the $\rho$-mesons. At energies much larger than the QCD mass scale the spectral functions are given by the continuum of quark-antiquark states. Analogously we expect that at the energy of the order of the QHD mass scale the spectral functions of the weak currents are dominated by the weak bosons. At energies much larger than the QHD mass scale there would be the continuum of haplon pairs.\\

In strong interaction physics the universality of the couplings of the $\rho$-mesons to the hadrons follows from the current algebra and the dominance of the matrix elements of the vector currents by the $\rho$-mesons. In the Standard Model the universality of the weak coupling constants is due to the gauge invariance. In our composite model of the weak bosons it follows from the algebra of the weak currents and the dominance of the weak currents by the weak bosons (ref.(3)).\\

In the absence of electromagnetism and the quark masses the three $\rho$-mesons are degenerate in mass. If the electromagnetic interaction is introduced, the neutral $\rho$-meson changes its mass due to a mixing with the photon. The mass
shift, caused by this mixing, can be calculated. It depends on a mixing parameter $\mu$, which is determined by the electric charge, the decay constant $F_\rho$ and the mass of the $\rho$-meson:\\

\begin{equation}
\mu \; =\; e \frac{F^{}_\rho}{M^{}_\rho}
\end{equation}\\

One finds for the mass difference between the charged and neutral  $\rho$-mesons:\\

\begin{equation}
\;{M(\rho}^0)^2\; - \;{M(\rho}^+)^2\ =\;{M\rho}^+)^2\;  \left(\frac{\mu \;^2}{1 - \mu \;^2}\right) \; .
\end{equation}\\

The decay constant is measured to about 220 MeV, which gives $\mu\approx 0.09$. The mass shift due to the mixing is about 3.1 MeV.\\

Analogously in QHD the three weak bosons are degenerate in mass in the absence of electromagnetism. If the electromagnetic interaction is introduced, the mass of the neutral boson increases due to the mixing with the photon. This mass shift has been measured in the LEP-experiments: $10.79 \pm 0.03$ GeV.\\

The mixing between the neutral weak boson and the photon it is caused by the dynamics, like the mixing between the photon and the neutral $\rho$-meson in QCD. It is described by a mixing parameter m, which is determined by the decay constant of the weak boson $F_W$, defined in analogy to the decay constant of the $\rho$-meson in QCD (ref.(3)):\\

\begin{equation}
\langle 0 \left| \frac{1}{2} \left( \overline{\alpha} \gamma^{}_\mu
\alpha - \overline{\beta} \gamma^{}_\mu \beta \right) \right| Z
\rangle \; =\; \varepsilon^{}_\mu M^{}_W F^{}_W \; .
\end{equation}\\

In the Standard Model the mixing is described by the weak mixing angle. In QHD the mixing is a dynamical phenomenon, and the mixing parameter m is given by the decay constant of the $W$-boson $F_W$ ( see eq. (4)):\\

\begin{equation}
m \; =\; e \frac{F^{}_W}{M^{}_W} \
\end{equation}\\

In the Standard Model the mixing parameter m is determined by the weak mixing angle ( ref.(3)):\\

\begin{equation}
\sin\theta^{}_w \; =\; m \
\end{equation}\\

The mass difference between the $Z$-boson  and the $W$-boson is given by the mixing parameter m and the $W$-mass:\\

\begin{equation}
M^2_Z - M^2_W \; =\; M^2_W \left(\frac{m^2}{1 - m^2}\right) \; .
\end{equation}\\

We shall use the following experimental values:

\begin{eqnarray}
M^{}_W & = & 80.4 ~{\rm GeV} \; , \nonumber \\
M^{}_Z & = & 91.9 ~{\rm GeV} \; , \nonumber \\
F^{}_W & = & 124 ~{\rm GeV} \; , \nonumber \\
\sin^2 \theta^{}_W & = & 0.2315 \; , \nonumber \\
\alpha & = & \frac{e^2}{4\pi} \; \approx \; \frac{1}{128.9} \; ,
\nonumber \\
e & = & 0.3122 \; , \nonumber \\
m & = & 0.482 \
\end{eqnarray}

In strong interaction physics the decay constant of the $\rho$-meson meson and the QCD mass parameter $\Lambda_c$ are proportional. The decay constant of the $\rho$-meson is measured to about 220 MeV. The QCD mass parameter $\Lambda_c$ has also been measured: $217 \pm 25$ MeV. Thus both parameters are about equal. We expect a similar connection between the decay constant of the weak boson and the QHD mass parameter $\Lambda_h$. If the QHD gauge group would be SU(3), the ratio of $\Lambda_h$ and $\Lambda_c$ would be given by the measured ratio of the decay constants:\\

\begin{equation}
\frac{\Lambda_h}{\Lambda_c}\;\approx\; \frac{F^{}_W}{F^{}_\rho}\;\simeq560.
\end{equation}\\

In that case $\Lambda_h$ would be 0.122 TeV. The value of $\Lambda_h$ depends on the gauge group, but it should be less than 0.5 TeV. Thus the mass scales of QCD and QHD differ by about three orders of magnitude.\\

If the weak bosons consist of the two haplons $\alpha$ and $\beta$, there must exist a second neutral weak boson $X$, which is an isoscalar and has the internal structure:\\

\begin{eqnarray}
X & = & \frac{1}{\sqrt{2}} \left( \overline{\alpha} \alpha +
\overline{\beta} \beta \right) \; .
\end{eqnarray}\\

This boson is not present in the Standard Model - its mass must be much larger than the mass of the $Z$-boson. The present lower limit on the mass of the $X$-boson is about 0.8 TeV (see also ref.(7)).\\

In strong interaction physics the mass of the $\rho$-meson and of the $\omega$-meson are nearly the same. One would expect that the mass of the $X$--boson is about the same as the mass of the $Z$-boson, but this is excluded by the experiments. The fact that the $X$--boson, if it exists, is much heavier than the $Z$-boson, might be related to the QHD analogy of the gluonic anomaly of QCD. The latter implies that the mass of the $\eta^{\prime}$-meson is different from zero in the chiral limit, while the masses of the $\pi$-mesons and of the $\eta$-meson vanish. In QHD the isospin singlet axial vector current also has an anomaly, and this might be the reason why the $X$-boson is very heavy. The theory of QHD is a confining chiral gauge theory, and low mass pseudoscalar bosons do not exist. The anomalous divergence of the singlet axial vector current might increase the mass of the $X$-boson. But details about the dynamics of chiral gauge theories are not yet known. For our further discussion we shall assume a mass of 0.8 TeV for the $X$-boson.\\

The $X$-boson would couple to the leptons and quarks with about the same strength as the $Z$-boson. Thus it can easily be produced at the LHC by quark-antiquark-fusion. The cross section for the production of $Z$-bosons at the LHC is estimated to about 60 nb. If the $X$-boson has a mass of 0.8 TeV, we can estimate the cross section for its production at the LHC. It should be about 0.8 nb.\\

An important decay mode of the $X$-boson is the decay into lepton pairs, e.g. into muon pairs. The partial width for this decay can be estimated by comparing it with the decay of a charged weak boson into a muon and the muon antineutrino, which has a partial width of about 220 MeV. Using this result, we can calculate the partial width for the leptonic decay of the $X$-boson with a mass of 0.8 TeV:\\

\begin{equation}
\Gamma(X\rightarrow \mu^+ \mu^-)\approx\,2.24\,GeV.
\end{equation}\\

The $X$-boson will decay primarily into lepton pairs and quark pairs.
We expect the following relations to hold between the
branching fractions of the various decays:\\

\begin{eqnarray}
&&Br(X\rightarrow e^+ e^-)\cong Br(X\rightarrow \nu_e \nu_e) \nonumber\\
&&Br(X\rightarrow \nu_e \nu_e)\cong Br(X\rightarrow \nu_\mu\nu_\mu)\cong Br(X\rightarrow \nu_\tau\nu_\tau) \nonumber\\
&&Br(X\rightarrow e^+ e^-)\cong Br(X\rightarrow \mu^+ \mu^-)\cong Br(X\rightarrow \tau^+ \tau^-) \nonumber\\
&&3Br(X\rightarrow e^+ e^-)\cong Br(X\rightarrow \overline{u}u)\cong Br(X\rightarrow \overline{d}d)\nonumber\\
&&Br(X\rightarrow \overline{u}u)\cong Br(X\rightarrow \overline{c}c)\cong Br(X\rightarrow \overline{t}t) \nonumber\\
&&Br(X\rightarrow \overline{d}d)\cong Br(X\rightarrow \overline{s}s)\cong Br(X\rightarrow \overline{b}b)
\end{eqnarray}\\

The $X$-boson can also decay into weak bosons. We expect:\\

\begin{equation}
\Gamma(X\rightarrow W^+W^-)\cong\Gamma(X\rightarrow Z Z)\simeq\Gamma(X\rightarrow \mu^+ \mu^-)
\end{equation}\\

For the decays of the $X$ - boson into three and four weak bosons
the following relations for the partial widths are expected:\\

\begin{eqnarray}
&&\Gamma(X\rightarrow W^+W^-Z)\cong\Gamma(X\rightarrow ZZZ)\nonumber\\
&&\Gamma(X\rightarrow W^+W^-W^+W^-)\cong\Gamma(X\rightarrow W^+W^-ZZ)\nonumber\\
&&\Gamma(X\rightarrow W^+W^-W^+W^-)\cong\Gamma(X\rightarrow ZZZZ)
\end{eqnarray}\\

We introduce the parameters a and b:
\begin{eqnarray}
&&\Gamma(X\rightarrow W^+W^-Z)= a \cdot \Gamma(X\rightarrow \mu^+ \mu^-)\nonumber\\
&&\Gamma(X\rightarrow W^+W^-W^+W^-)=b \cdot \Gamma(X\rightarrow \mu^+ \mu^-)
\end{eqnarray}\\

 The parameters "a" and "b" are expected to be smaller than one. As an example we set a=0.5, b=0.3 and estimate the total width of the $X$-boson. There are three decay channels for the charged leptons, three channels for the neutrinos and 18 channels for the quark-antiquark pairs (including the color degree of freedom). The decays into weak bosons are added, using the parameters above. Other decays of the $X$ - boson are expected to be small and are neglected. Then we find for the total width:\\

\begin{equation}
\Gamma(X\rightarrow all)\approx 63 ~ GeV.
\end{equation}\\

The $Z$-boson has a width of 2.5 GeV - thus the width of the $X$-boson is about 25 times larger.\\ 

The best way to observe the $X$ - boson in the collisions at the LHC is to find the decays into muon pairs and into electron-positron pairs. Once it has been found, one can search for the decays into quark-antiquark pairs. Two narrow quark jets should be observed with an invariant mass given be the mass of the $X$-boson.\\

The QHD mass scale $\Lambda_h$ is three orders of magnitude larger than the QCD mass scale $\Lambda_c$. In strong interaction physics complexities arise at the energy of 1 GeV and above. Analogously there should be complexities due to the QHD dynamics at the energy of 1 TeV and above. In strong interaction physics there exist excited states of the vector mesons. Analogously we expectnexcited states of the weak bosons, with masses of the order of 1 TeV and higher. These states can have the total angular momentum 1,2,3 etc. The excited state of a charged weak boson can decay into a charged weak boson and a $Z$-boson or a photon. The excited state of a $Z$-boson will decay mainly into two weak bosons. Decays of the excited weak bosons into quark pairs or leptons pairs are expected to be suppressed.\\

The weak bosons couple universally to the leptons and quarks, as the $\rho$-mesons to the nucleons. Inside the $\rho$-meson are the same quarks as inside the nucleons, and this leads to the universality of the coupling parameters. Analogously we expect form the universality of the weak coupling parameters that inside a lepton and quark the haplons $\alpha$ and $\beta$ are also present. A bound state model of the weak bosons requires that the leptons and quarks are also composite systems.\\

The simplest model of composite leptons and quarks  is the one discussed in ref.(4). Besides the fermions $\alpha$ and $\beta$ four scalar haplons are needed, one scalar for the leptons,
denoted by l, and three scalars for the three colors of the quarks, denoted by r, g and b:\\

\begin{equation}
h(fermion) = \left( \begin{array}{l}
\alpha\\
\beta\\
\end{array} \right) \
\end{equation}

\begin{equation}
h(scalar) = (l,\
r,\
g,\
b)\
\end{equation}

Both the fermions and the scalars transform according to the fundamental representation of the QHD gauge group. Thus bound states of the fermions and the scalars exist - the lowest states would be the observed leptons and quarks. The lowest leptons have the following internal structure:\\

\begin{eqnarray}
{\nu} =( {\alpha} l)\nonumber \\
e^-=( {\beta} l) \nonumber \\
\end{eqnarray}

The structure of the up and down quarks (with red color) is given by:\\

\begin{eqnarray}
u =( {\alpha} r) \nonumber \\
d =( {\beta} r) \nonumber \\
\end{eqnarray}

In such a model the first generation of leptons and quarks would describe the ground states of the fermion-scalar bound states, the second and third generation must be dynamical excitations - the electron is the ground state of the charged leptons, the muon and the tau-lepton are excitations. Likewise the u-quark is the ground state of the up-quarks, the c-quark and the t-quarks are excitations. Compared to the QHD mass scale $\Lambda_h$ the masses of the observed leptons and quarks are essentially zero. The number of nearly massless bound states could be related to the rank of the QHD gauge group. Three generations might be obtained, if the gauge group is $SU(3)$.\\

The cross section in proton-proton collisions for exciting the $QHD$ degrees of freedom can be estimated as follows. The size of the proton is about 1 Fermi. The inelastic cross section is about 60 mb. The size of a quark is about 0.001 Fermil. Thus the cross section for exciting the $QHD$ degrees of freedom in quark-quark-collisions is about 60 nb. In the proton there are three quarks and many gluons. We estimate the cross section for exciting the $QHD$ degrees of freedom to about 600 nb.\\  

For the LHC of particular interest is the scattering of a quark and the corresponding antiquark. If these quarks collide, the scalars inside the quark or antiquark collide and can form a resonance. This resonance, formed e.g. by the collision of red scalar and its antiparticle, can decay into a leptonic scalar and its antiparticle. This scalar will form together with the fermion a lepton, e.g. a muon. Thus the quark and antiquark disappear, and a muon and its antiparticle are produced, with an invariant mass, given by the mass of the resonance. Likewise an electron and a positron can be produced, or a tau-lepton and its antiparticle. We expect that the masses of these resonances start at about 1 TeV - these resonances can be observed at the LHC.\\

We expect that the first signal of the new substructure of the weak bosons is the discovery of the $X$-boson at the Large Hadron Collider at CERN.\\

{\end{document} }}
\begin{thebibliography}{99}

\bibitem{4}  H. Fritzsch and G. Mandelbaum, Phys. Lett. B102 (1981) 319; \\
    Phys. Lett. B 109 (1982) 224; \\ X. Calmet and H. Fritzsch, Phys. Lett. B496 (2000), 161.

\bibitem{5}  R. Barbieri, R. Mohapatra and A. Masiero, Phys. Lett. B 105
(1981) 369.

\bibitem{6}  H. Fritzsch. R. Kogerler and  D. Schildknecht, Phys. Lett. B 114 (1982)
157.

\bibitem{7}  L. F. Abbott and E. Farhi, Phys. Lett. B 101, 69
(1981).

\bibitem{8} T. Kugo, S. Uehara and T. Yanagida, Phys. Lett. B 147, 321
(1984).

\bibitem{9} S. Uehara and T. Yanagida, Phys. Lett. B 165, 94
(1985).

\bibitem{10}  U. Baur and K. H. Schwarzer, Phys. Lett. B 180, 163
(1986).

\end{thebibliography}
